\documentclass[11pt]{article}

\usepackage{lipsum} 

\usepackage{bm}
\usepackage{floatrow} 
\pdfoutput=1
\usepackage{amsmath,amsfonts,amsthm}
\usepackage{esdiff}  
\usepackage{booktabs}  
\usepackage{url}  
\usepackage{hyperref}  

\usepackage[tableposition=top]{caption}

\usepackage{cleveref}  
	\crefname{equation}{equation}{equations}
	\crefname{figure}{figure}{figures}	
	\crefname{table}{table}{tables}
\usepackage[skip=1.5pt,font=small]{caption}

\usepackage{bbm}

\usepackage[usenames,dvipsnames,svgnames,table]{xcolor}

\usepackage[export]{adjustbox}
\usepackage{graphicx}

\usepackage{cite}

\usepackage{tikz}  

\usepackage[sc]{mathpazo} 
\usepackage[T1]{fontenc} 
\usepackage{microtype} 

\usepackage{multicol} 
\usepackage[margin={1cm,1.5cm}]{geometry}
\usepackage{booktabs} 
\usepackage{float} 
\usepackage{hyperref} 

\usepackage{lettrine} 
\usepackage{paralist} 

\usepackage{abstract} 

\usepackage{soul}
\usepackage{titlesec} 
\renewcommand\thesection{\Roman{section}} 
\renewcommand\thesubsection{\Alph{subsection}} 
\titleformat{\section}[block]{\large\scshape\centering\bfseries}{\thesection.}{1em}{} 

\titleformat{\subsection}[block]{\scshape\centering}{\thesubsection.}{1em}{} 

\usepackage{fancyhdr} 
\DeclareCaptionFormat{myformat}{#1#2#3\hrulefill}
\usepackage{float}
\floatstyle{plaintop}
\restylefloat{table}

\usepackage{authblk}
\makeatletter

\title{\vspace{-15mm}\fontsize{15pt}{15pt}\selectfont\textbf{Identifying the measurements required to estimate rates of COVID-19 transmission, infection, and detection, using variational data assimilation}}  
\author[1,2]{Eve Armstrong\thanks{evearmstrong.physics@gmail.com}}
\author[3]{Manuela Runge}
\author[3]{Jaline Gerardin}
\affil[1]{Department of Physics, New York Institute of Technology, New York, NY 10023, USA}
\affil[2]{Department of Astrophysics, American Museum of Natural History, New York, NY 10024, USA}
\affil[3]{Department of Preventive Medicine, Northwestern University, Chicago IL, 60611}

\par
\date{\today}
\begin{document}
\maketitle 
  
\begin{abstract}
We demonstrate the ability of statistical data assimilation to identify the measurements required for accurate state and parameter estimation in an epidemiological model for the novel coronavirus disease COVID-19.  Our context is an effort to inform policy regarding social behavior, to mitigate strain on hospital capacity.  The model unknowns are taken to be: the time-varying transmission rate, the fraction of exposed cases that require hospitalization, and the time-varying detection probabilities of new asymptomatic and symptomatic cases.  In simulations, we obtain accurate estimates of undetected (that is, unmeasured) infectious populations, by measuring the detected cases together with the recovered and dead - and without assumed knowledge of the detection rates. Given a noiseless measurement of the recovered population, excellent estimates of all quantities are obtained using a temporal baseline of 101 days, with the exception of the time-varying transmission rate at times prior to the implementation of social distancing.  With low noise added to the recovered population, accurate state estimates require a lengthening of the temporal baseline of measurements.  Estimates of all parameters are sensitive to the contamination, highlighting the need for accurate and uniform methods of reporting.  The aim of this paper is to exemplify the power of SDA to determine what properties of measurements will yield estimates of unknown parameters to a desired precision, in a model with the complexity required to capture important features of the COVID-19 pandemic.
\end{abstract}
\maketitle
\begin{multicols}{2}
\section{INTRODUCTION}
The coronavirus disease 2019 (COVID-19) is burdening health care systems worldwide, threatening physical and psychological health, and devastating the global economy. Individual countries and states are tasked with balancing population-level mitigation measures with maintaining economic activity. Mathematical modeling has been used to aid policymakers' plans for hospital capacity needs, and to understand the minimum criteria for effective contact tracing~\cite{team_forecasting_2020}.  Both state-level decision-making and accurate modeling benefit from quality surveillance data.  Insufficient insufficient testing capacity, however, especially at the beginning of the epidemic in the United States, and other data reporting issues have meant that surveillance data on COVID-19 is biased and incomplete~\cite{needGoodData2020,weinberger2020estimating,li2020substantial}. Models intended to guide intervention policy must be able to handle imperfect data.

Within this context, we seek a means to quantify what data must be recorded in order to estimate specific unknown quantities in an epidemiological model of COVID-19 transmission.  These unknown quantities are: i) the transmission rate, ii) the fraction of the exposed population that acquires symptoms sufficiently severe to require hospitalization, and iii) time-varying detection probabilities of asymptomatic and symptomatic cases.  In this paper, we demonstrate the ability of statistical data assimilation (SDA) to quantify the accuracy to which these parameters can be estimated, given certain properties of the data including noise level.  

SDA is an inverse formulation~\cite{tarantola2005inverse}: a machine learning approach designed to optimally combine a model with data.  Invented for numerical weather prediction~\cite{kimura2002numerical,kalnay2003atmospheric,evensen2009data,betts2010practical,whartenby2013number,an2017estimating}, and more recently applied to biological neuron models~\cite{schiff2009kalman,toth2011dynamical,kostuk2012dynamical,hamilton2013real,meliza2014estimating,nogaret2016automatic,armstrong2020statistical}, SDA offers a systematic means to identify the measurements required to estimate unknown model parameters to a desired precision.  

Data assimilation has been presented as a means for general epidemiological forecasting~\cite{bettencourt2007towards}, and one work has examined variational data assimilation specifically - the method we employ in this paper - for estimating parameters in epidemiological models~\cite{rhodes2009variational}.  Related Bayesian frameworks for estimating unknown properties of epidemiological models have also been explored~\cite{cobb2014bayesian,bettencourt2008real}.  To date, there have been two employments of SDA for COVID-19 specifically.  Ref~\cite{sesterhenn2020adjoint} used a simple SIR (susceptible/infected/recovered) model, and Ref~\cite{nadler2020epidemiological} expanded the SIR model to include a compartment of patients in treatment.  

Two features of our work distinguish this paper as novel.  First, we expand the model in terms of the number of compartments.  The aim here is to capture key features of COVID-19 such that the model structure is relevant for questions from policymakers on containing the pandemic.  These features are: i) asymptomatic, presymptomatic, and symptomatic populations, ii) undetected and detected cases, and iii) two hospitalized populations: those who do and do not require critical care.  For our motivations for these choices, see \textit{Model}.  Second, we employ SDA for the specific purpose of examining the sensitivity of estimates of time-varying parameters to various properties of the measurements, including the degree of noise (or error) added.  Moreover, we aim to demonstrate the power and versatility of the SDA technique to explore what is required of measurements to complete a model with a dimension sufficiently high to capture the policy-relevant complexities of COVID-19 transmission and containment - an examination that has not previously been done.

To this end, we sought to estimate the parameters noted above, using simulated data representing a metropolitan-area population loosely based on New York City.  We examined the sensitivity of estimations to: i) the subpopulations that were sampled, ii) the temporal baseline of sampling, and iii) uncertainty in the sampling.  

Results using simulated data are threefold.  First, reasonable estimations of time-varying detection probabilities require the reporting of new detected cases (asymptomatic and symptomatic), dead, and recovered.  Second, given noiseless measurements, a temporal baseline of 101 days is sufficient for the SDA procedure to capture the general trends in the evolution of the model populations, the detection probabilities, and the time-varying transmission rate following the implementation of social distancing.  Importantly, the information contained in the measured \textit{detected} populations propagates successfully to the estimation of the numbers of severe \textit{undetected} cases.  Third, the state evolution - and importantly the populations requiring inpatient care - tolerates low ($\sim$ five percent) noise, given a doubling of the temporal baseline of measurements; the parameter estimates do not tolerate this contamination.  

Finally, we discuss necessary modifications prior to testing with real data, including lowering the sensitivity of parameter estimates to noise in data.
 
\section{MODEL}
\begin{figure}[H]
  \includegraphics[width=0.99\textwidth,valign=t]{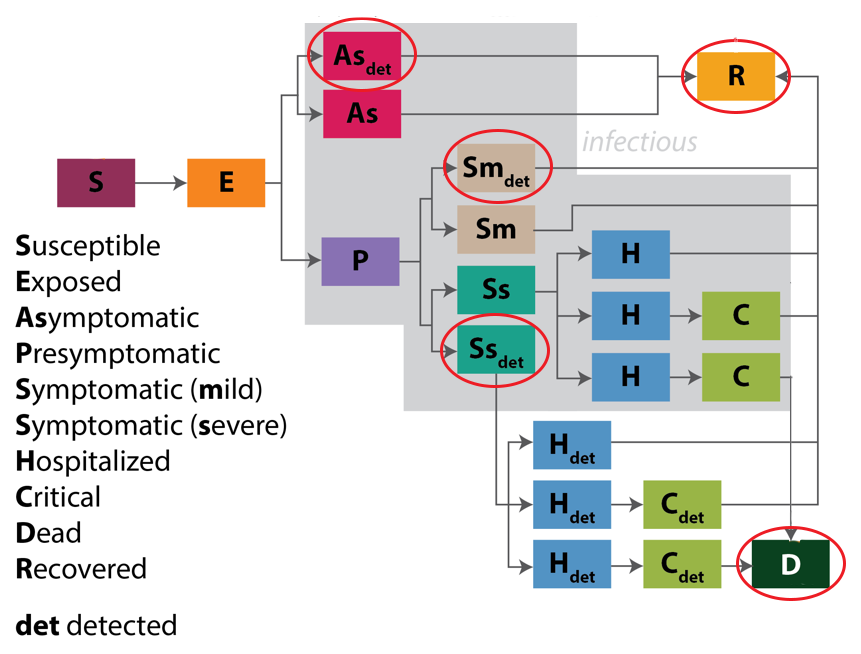}
 \caption{\textbf{Schematic of the model}.  Each rectangle represents a population.  Note the distinction of asymptomatic cases, undetected cases, and the two tiers of hospitalized care: $H$ and $C$.  The aim of including this degree of resolution is to inform policy on social behavior so as to minimize strain on hospital capacity.  The red ovals indicate the variables that correspond to measured quantities in the inference experiments.}
 \label{fig:schematic}
\end{figure}

The model is written in 22 state variables, each representing a subpopulation of people; the total population is conserved.  Figure~\ref{fig:schematic} shows a schematic of the model structure.  Each member of a Population $S$ that becomes Exposed ($E$) ultimately reaches either the Recovered ($R$) or Dead ($D$) state. \textit{Absent additive noise, the model is deterministic.} Five variables correspond to measured quantities in the inference experiments.

As noted, the model is written with the aim to inform policy on social behavior and contact tracing so as to avoid exceeding hospital capacity. To this end, the model resolves asymptomatic-versus-symptomatic cases, undetected-versus-detected cases, and the two tiers of hospital needs: the general (inpatient, non-intensive care unit (ICU)) $H$ versus the critical care (ICU) $C$ populations. 
\noindent
The resolution of asymptomatic versus symptomatic cases was motivated by an interest in what interventions are necessary to control the epidemic.  For example, is it sufficient to focus only on symptomatic individuals, or must we also target and address asymptomatic individuals who may not even realize they are infected? 

The detected and undetected populations exist for two reasons.  First, we seek to account for underreporting of cases and deaths.  Second, we desire a model structure that can simulate the impact of increasing detection rates on disease transmission, including the impact of contact tracing.  Thus the model was structured from the beginning so that we might examine the effects of interventions that were imposed later on.  The ultimate aim here is to inform policy on the requirements for containing the epidemic.

We included both $H$ and $C$ populations because hospital inpatient and ICU bed capacities are the key health system metrics that we aim to avoid straining.  Any policy that we consider must include predictions on inpatient and ICU bed needs.  Preparing for those needs is a key response if or when the epidemic grows uncontrolled.

For details of the model, including the reaction equations and descriptions of all state variables and parameters, see \textit{Appendix A}.
\section{METHOD}
\subsection{\textbf{General inference formulation}}
SDA is an inference procedure, or a type of machine learning, in which a model dynamical system is assumed to underlie any measured quantities.  This model $\bm{F}$ can be written as a set of \textit{D} ordinary differential equations that evolve in some parameterization $t$ as:
\begin{align*}
  \diff{x_a(t)}{t} &= F_a(\bm{x}(t),\bm{p}(t)); \hspace{1em} a =1,2,\ldots,D,
\end{align*}
\noindent
where the components $x_a$ of the vector \textbf{x} are the model state variables, and unknown parameters to be estimated are contained in $\bm{p}(t)$.  A subset \textit{L} of the \textit{D} state variables is associated with measured quantities.  One seeks to estimate the $p$ unknown parameters and the evolution of all state variables that is consistent with the $L$ measurements.

A prerequisite for estimation using real data is the design of simulated experiments, wherein the true values of parameters are known. In addition to providing a consistency check, simulated experiments offer the opportunity to ascertain \textit{which} and \textit{how few} experimental measurements, in principle, are necessary and sufficient to complete a model.  

\subsection{\textbf{Optimization framework}}

SDA can be formulated as an optimization, wherein a cost function is extremized.  We take this approach, and write the cost function in two terms: 1) one term representing the difference between state estimate and measurement (measurement error), and 2) a term representing model error.  It will be shown below in this Section that treating the model error as finite offers a means to identify whether a solution has been found within a particular region of parameter space.  This is a non-trivial problem, as any nonlinear model will render the cost function non-convex.  We search the surface of the cost function via the variational method, and we employ a method of annealing to identify a lowest minumum - a procedure that has been referred to loosely in the literature as variational annealing (VA). 

The cost function $A_0$ used in this paper is written as:
\begin{align}
\label{eq:costfunction}
  A_0(\bm{x}(n),\bm{p}) &= \sum_{j=1}^J\sum_{l=1}^{L} \frac{R^{l}_{m}}{2}(y_l(n) - x_{l}(n))^2 \notag\\
&+ \sum_{n=1}^{N-1}\sum_{a=1}^{D} \frac{R^{a}_{f}}{2}\left(x_a(n+1) -
f_a(\bm{x}(n),\bm{p}(n))\right)^2.
\end{align}
\noindent
One seeks the path $\bm{X}^0 = {\bm{x}(0),...,\bm{x}(N),\bm{p}(0),...\bm{p}(N)}$ in state space on which $A_0$ attains a minimum value\footnote{It may interest the reader that one can derive this cost function by considering the classical physical Action on a path in a state space, where the path of lowest Action corresponds to the correct solution~\cite{abarbanel2013predicting}}.  Note that Equation~\ref{eq:costfunction} is shorthand; for the full form, see \textit{Appendix A} of Ref~\cite{armstrong2020statistical}.  For a derivation - beginning with the physical Action of a particle in state space - see Ref~\cite{abarbanel2013predicting}.

The first squared term of Equation~\ref{eq:costfunction} governs the transfer of information from measurements $y_l$ to model states $x_l$.  The summation on \textit{j} runs over all discretized timepoints $J$ at which measurements are made, which may be a subset of all integrated model timepoints.  The summation on \textit{l} is taken over all \textit{L} measured quantities. 

The second squared term of Equation~\ref{eq:costfunction} incorporates the model evolution of all \textit{D} state variables $x_a$.  The term $f_a(\bm{x}(n))$ is defined, for discretization, as: $\frac{1}{2} [F_a(\bm{x}(n)) + F_a(\bm{x}(n+1))]$.  The outer sum on \textit{n} is taken over all discretized timepoints of the model equations of motion.  The sum on \textit{a} is taken over all \textit{D} state variables.

$R_m$ and $R_f$ are inverse covariance matrices for the measurement and model errors, respectively.  We take each matrix to be diagonal and treat them as relative weighting terms, whose utility will be described below in this Section. 

The procedure searches a $(D \,(N+1)+ p \,(N+1))$-dimensional state space, where \textit{D} is the number of state variables, \textit{N} is the number of discretized steps, and \textit{p} is the number of unknown parameters.  To perform simulated experiments, the equations of motion are integrated forward to yield simulated data, and the VA procedure is challenged to infer the parameters and the evolution of all state variables - measured and unmeasured - that generated the simulated data.

This specific formulation has been tested with chaotic models~\cite{abarbanel2011dynamical,ye2014estimating,rey2014accurate,ye2015improved}, and used to estimate parameters in models of biological neurons~\cite{toth2011dynamical,kostuk2012dynamical,meliza2014estimating,kadakia2016nonlinear,wang2016data,armstrong2020statistical}, as well as astrophysical scenarios~\cite{armstrong2017optimization}. 

\subsection{\textbf{Annealing to identify a solution on a non-convex cost function surface}}

Our model is nonlinear, and thus the cost function surface is non-convex.  For this reason, we iterate - or anneal - in terms of the ratio of model and measurement error, with the aim to gradually freeze out a lowest minimum.  This procedure was introduced in Ref~\cite{ye2015systematic}, and has since been used in combination with variational optimization on nonlinear models in Refs~\cite{an2017estimating,armstrong2020statistical,kadakia2016nonlinear,armstrong2017optimization} above.  The annealing works as follows.

We first define the coefficient of measurement error $R_m$ to be 1.0, and write the coefficient of model error $R_f$ as: $R_f = R_{f,0}\alpha^{\beta}$, where $R_{f,0}$ is a small number near zero, $\alpha$ is a small number greater than 1.0, and $\beta$ is initialized at zero.  Parameter $\beta$ is our annealing parameter.  For the case in which $\beta = 0$, relatively free from model constraints the cost function surface is smooth and there exists one minimum of the variational problem that is consistent with the measurements.  We obtain an estimate of that minimum.  Then we increase the weight of the model term slightly, via an integer increment in $\beta$, and recalculate the cost.  We do this recursively, toward the deterministic limit of $R_f \gg R_m$.  The aim is to remain sufficiently near to the lowest minimum to not become trapped in a local minimum as the surface becomes resolved.  We will show in \textit{Results} that a plot of the cost as a function of $\beta$ reveals whether a solution has been found that is consistent with both measurements and model.  

\section{THE EXPERIMENTS}
\subsection{\textbf{Simulated experiments}}
We based our simulated locality loosely on New York City, with a population of 9 million. For simplicity, we assume a closed population.  Simulations ran from an initial time $t_0$ of four days prior to 2020 March 1, the date of the first reported COVID-19 case in New York City~\cite{firstInNYC}.  At time $t_0$, there existed one detected symptomatic case within the population of 9 million.  In addition to that one initial detected case, we took as our initial conditions on the populations to be: 50 undetected asymptomatics, 10 undetected mild symptomatics, and one undetected severe symptomatic.

We chose five quantities as unknown parameters to be estimated (Table~\ref{tableUnknownPs}): 1) the time-varying transmission rate $K_i$(t); 2) the detection probability of mild symptomatic cases $d_{Sym}(t)$, 3) the detection probability of severe symptomatic cases $d_{Sys}(t)$, 4) the fraction of cases that become symptomatic $f_{sympt}$, and 5) the fraction of symptomatic
cases that become severe enough to require hospitalization $f_{severe}$.  Here we summarize the key features that we sought to capture in modeling these parameters; for their mathematical formulatons, see \textit{Appendix B}.  
\end{multicols}
\setlength{\tabcolsep}{5pt}
\begin{table}[H]
\small
\centering
\begin{tabular}{l l} \toprule
 \textit{Parameter} & \textit{Description}\\\midrule
 $K_i(t)$ & Time-varying transmission rate\\
 $d_{Sym}(t)$ & Time-varying detection probability of mild symptomatics\\
 $d_{Sys}(t)$ & Time-varying detection probability of symptomatics requiring hospitalization\\
 $f_{sympt}$ & Fraction of positive cases that produce symptoms \\
 $f_{severe}$ & Fraction of symptomatics that are severe\\\bottomrule
 \end{tabular}
\caption{\textbf{Unknown parameters to be estimated.}  $K_i$, $d_{Sym}$, and $d_{Sys}$ are taken to be time-varying.  Parameters $f_{sympt}$ and $f_{severe}$ are constant numbers, as they are assumed to reflect an intrinsic property of the disease.  The detection probability of asymptomatic cases is taken to be known and zero.} 
\label{tableUnknownPs}
\end{table}
\begin{multicols}{2}
\noindent

The transmission rate $K_i$ (often referred to as the effective contact rate) in a given population for a given infectious disease is measured in effective contacts per unit time.  This may be expressed as the total contact rate multiplied by the risk of infection, given contact between an infectious and a susceptible individual.  The contact rate, in turn, can be impacted by amendments to social behavior\footnote{The reproduction number $R_0$, in the simplest SIR form, can be written as the effective contact rate divided by the recovery rate.  In practice, $R_0$ is a challenge to infer~\cite{thompson2019improved,cori2013new,bettencourt2008real,wallinga2004different}.
}.

As a first step in applying SDA to a high-dimensional epidemiological model, we chose to condense the significance of $K_i$ into a relatively simple mathematical form.  We assumed that $K_i$ was constant prior to the implementation of a social-distancing mandate, which then effected a rapid transition of $K_i$ to a lower constant value.  Specifically, we modeled $K_i$ as a smooth approximation to a Heaviside function that begins its decline on March 22, the date that the stay-at-home order took effect in New York City~\cite{PAUSEorder}: 25 days after time $t_0$.  For further simplicity, we took $K_i$ to reflect a single implementation of a social distancing protocol, and adherence to that protocol throughout the remaining temporal baseline of estimation.  

Detection rates impact the sizes of the subpopulations entering hospitals, and their values are highly uncertain~\cite{weinberger2020estimating,li2020substantial}.  Thus we took these quantities to be unknown, and - as detection methods will evolve - time-varying.  We also optimistically assumed that the methods will improve, and thus we described them as increasing functions of time.  We used smoothly-varying forms, the first linear and the second quadratic, to preclude symmetries in the model equations.  Meanwhile, we took the detection probability for asymptomatic cases ($d_{As}$) to be known and zero, a reasonable reflection of the current state of testing.  

Finally, we assigned as unknowns the fraction of cases that become symptomatic ($f_{sympt}$) and fraction of symptomatic cases that become sufficiently severe to require hospitalization ($f_{severe}$), as these fractions possess high uncertainties (Refs~\cite{fSympt} and \cite{salje2020estimating}, respectively).  As they reflect an intrinsic property of the disease, we took them to be constants.  All other model parameters were taken
to be known and constant (\textit{Appendix A}); however, the values of many other model parameters also possess significant uncertainties given the reported data, including, for example, the fraction of those hospitalized that require ICU care.  Future VA experiments can treat these quantities as unknowns as well.
\begin{figure}[H]
  \includegraphics[width=0.99\textwidth,valign=t]{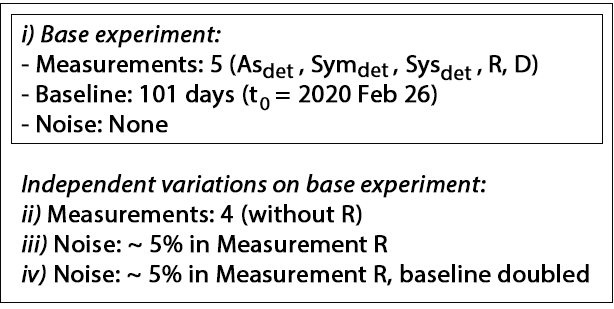}
 \caption{\textbf{Schematic of the four simulated experiments.}}
 \label{fig:experLayout}
\end{figure}
The simulated experiments are summarized in the schematic of Figure~\ref{fig:experLayout}.  They were designed to probe the effects upon estimations of three considerations: a) the number of measured subpopulations, b) the temporal baseline of measurements, and c) contamination of measurements by noise.  To this end, we designed a \lq\lq base\rq\rq\ experiment sufficient to yield an excellent solution, and then four variations on this experiment. 

The base experiment ($i$ in Figure~\ref{fig:experLayout}) possesses the following features: a) five measured populations: detected asymptomatic $As_{det}$, detected mild symptomatic $Sym_{det}$, detected severe symptomatic $Sys_{det}$, Recovered $R$, and Dead $D$; b) a temporal baseline of 101 days, beginning on 2020 February 26; c) no noise in measurements. 

The three variations on this basic experiment ($ii$ through $iv$ in Figure~\ref{fig:experLayout}), incorporate the following independent changes.  In Experiment $ii$, the $R$ population is not measured - an example designed to reflect the current situation in some localities (e.g. Refs~\cite{weinberger2020estimating,li2020substantial}).  

Experiment $iii$ includes a $\sim$ five percent noise level (for the form of additive noise, see \textit{Appendix C}) in the simulated $R$ data, and Experiment $iv$ includes that noise level in addition to a doubled temporal baseline. 

For each experiment, twenty independent calculations were initiated in parallel searches, each with a randomly-generated set of initial conditions on state variable and parameter values.  For technical details of all experimental designs and implementation, see \textit{Appendix C}.

\section{RESULT}
\subsection{\textbf{General findings}}

The salient results for the simulated experiments $i$ through $iv$ are as follows: 
\begin{enumerate}[i]
  \item (base experiment): Excellent estimate of all - measured and unmeasured - state variables, and all parameters except for $K_i$(t) at times prior to the onset of social distancing;
  \item (absent a measurement of Population $R$): Poor estimate of all quantities;
  \item ($\sim$ 5\% additive noise in $R$): Poor estimates of all quantities;
  \item ($\sim$ 5\% additive noise in $R$, with a doubled baseline of 201 days): Estimates of state evolution are robust to noise, while parameter estimates are sensitive to noise.
\end{enumerate}
Figures of the estimated time evolution of state variables and time-varying parameters are shown in their respective subsections, and the estimates of the static parameters are listed in Table~\ref{table:Pestimates}.

\subsection{\textbf{Base Experiment $\mathbf{i}$}}

The base experiment that employed five noiseless measured populations over 101 days yielded an excellent solution in terms of model evolution and parameter estimates.  Prior to examining the solution, we shall first show the cost function versus the annealing parameter $\beta$, as this distribution can serve as a tool for assessing the significance of a solution.
\setlength{\tabcolsep}{5pt}
\begin{table}[H]
\small
\centering
\begin{tabular}{l | l l | l l } \toprule
 \textit{Experiment} & $f_{sympt}$ & (true: 0.6) & $f_{severe}$ & (true: 0.07)\\\midrule
 & \textit{Mean} & \textit{Variance} & \textit{Mean} & \textit{Variance} \\\midrule
 $i$ & 0.59 & $2\times10^{-4}$ & 0.07 & $4\times10^{-6}$\\
 $ii$ & -- & & & \\
 $iii$ & -- & & & \\
 $iv$ & 0.39 & 0.8 & 0.19 & 0.2 \\\bottomrule
 \end{tabular}
\caption{\textbf{Estimates of static parameters $\bm{f_{sympt}}$ and $\bm{f_{severe}}$} over all simulated experiments.  For Experiments $i$ and $iv$, the reported numbers are taken from the annealing iteration with a value of parameter $\beta$ of 32 and 40, respectively: once the deterministic limit has been reached (see text).  For Experiment $ii$, an attempt was made to retrieve parameter estimates at $\beta=2$; that is: before the solution grows unstable exponentially (see Figure~\ref{fig:Action_nM5_D101_noNoise}).  See specific subsections for details of each experiment.} 
\label{table:Pestimates}
\end{table}

\begin{figure}[H]
  \includegraphics[width=0.8\textwidth,valign=t]{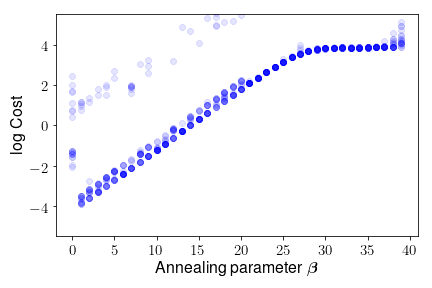}
\caption{\textbf{Cost function plotted at each annealing step $\bm{\beta}$ for the base experiment $\bm{i}$, for twenty paths in state space, where $\beta$ scales the rigidity of the imposed model constraint.}  At low $\beta$ the procedure endeavours to fit the measured variables to the simulated measurements.  As $\beta$ increases, the cost increases until it approaches a plateau (around $\beta = 30$), indicating that a solution has been found that is consistent with both measurements and model.}
 \label{fig:Action_nM6_D101_noNoise}
\end{figure}

\begin{figure}[H]
  \includegraphics[width=0.9\textwidth]{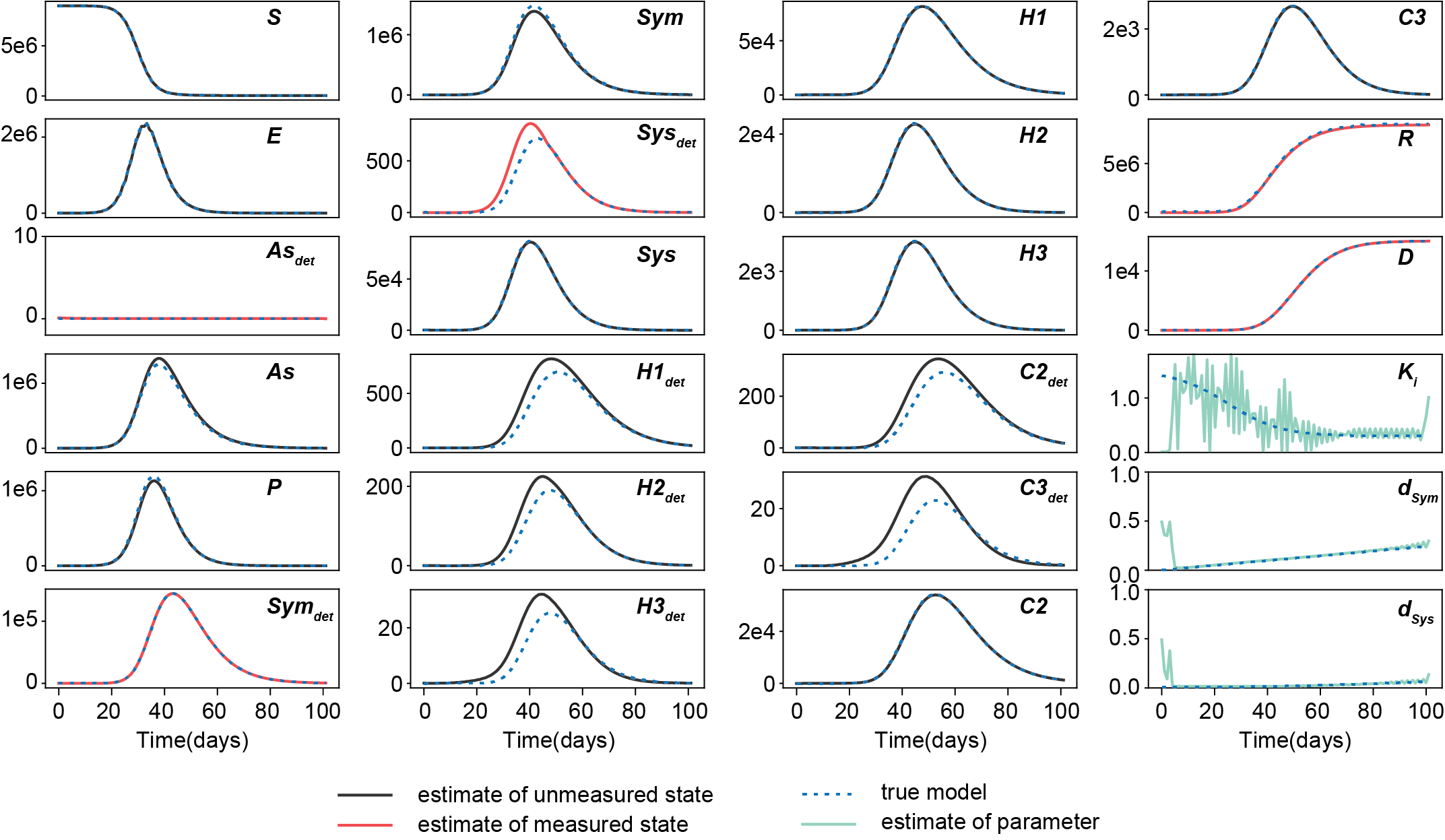}
\caption{\textbf{Estimates of the state - measured and unmeasured - variables, and the time-varying parameters $\bm{K_i}$, $\bm{d_{Sym}}$, and $\bm{d_{Sys}}$, for the base experiment $\bm{i}$.}  Excellent estimates are obtained of all states and parameters, except early values of $K_i$ prior to the implementation of social distancing; see text.  Results are taken at a value for annealing parameter $\beta$ of 32.}
 \label{fig:state_nM6_D101_noNoise}
\end{figure}

Figure~\ref{fig:Action_nM6_D101_noNoise} shows the evolution of the cost throughout annealing, for the ten distinct independent paths that were initiated; the x-axis shows the value of Annealing Parameter $\beta$, or: the increasing rigidity of the model constraint.  At the start of iterations, the cost function is mainly fitting the measurements to data, and its value begins to climb as the model penalty is gradually imposed.  If the procedure finds a solution that is consistent not only with the measurements, but also with the model, then the cost will plateau.  In Figure~\ref{fig:state_nM6_D101_noNoise}, we see this happen, around $\beta = 30$, with some scatter across paths.  The reported estimates in this Subsection are taken at a value of $\beta$ of 32: on the plateau.  The significance of this plateau will become clearer upon examining the contrasting case of Experiment $ii$.

We now examine the state and parameter estimates for the base experiment $i$. For all experiments, each solution shown is representative of the solution for all twenty paths.  Figure~\ref{fig:state_nM6_D101_noNoise} shows an excellent estimate of all state variables during the temporal window in which the measured variables were sampled. For consistency in illustrating the time evolution of all state variables, we use the state estimates for the Recovered ($R$) and Dead ($D$) populations, which are cumulative, rather than follow standard epidemiological practice of showing incident $R$ or $D$.  The time-varying parameters are also estimated well, excepting $K_i$(t) at times prior to its steep decline.  We noted no improvement in this estimate for $K_i$(t), following a tenfold increase in the temporal resolution of measurements (not shown).  The procedure does appear 
\end{multicols}
\begin{multicols}{2}
\noindent
to recognize that a fast transition in the value of $K_i$ occurred at early times, and that that value was previously higher.  It will be important to investigate the reason for this failure in the estimation of $K_i$ at early times, to rule out numerical issues involved with the quickly-changing derivative\footnote{As noted in \textit{Experiments}, we chose $K_i$ to reflect a rapid adherence to social distancing at Day 25 following time $t_0$, which then remained in place through to Day 101.  For the form of $K_i$, see \textit{Appendix B}.)}. 
\subsection{\textbf{Experiment $\mathbf{ii}$: no measurement of $\mathbf{R}$}}

\begin{figure}[H]
  \includegraphics[width=0.8\textwidth,valign=t]{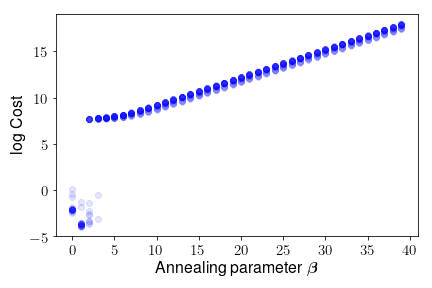} 
\caption{\textbf{Cost versus $\bm{\beta}$ for Experiment $\bm{ii}$: $\bm{R}$ is not measured.}  As $\beta$ increases, the cost increases indefinitely, indicating that no solution has been found that is consistent with both measurements and model dynamics.}
 \label{fig:Action_nM5_D101_noNoise}
\end{figure}

Figure~\ref{fig:Action_nM5_D101_noNoise} shows the cost as a function of annealing for the case with no measurement of Recovered Population $R$.  Without examining the estimates, we know from the Cost($\beta$) plot that no solution has been found that is consistent with both measurements and model: no plateau is reached.  Rather, as the model constraint strengthens, the cost increases exponentially.

\begin{figure}[H]
  \includegraphics[width=0.9\textwidth]{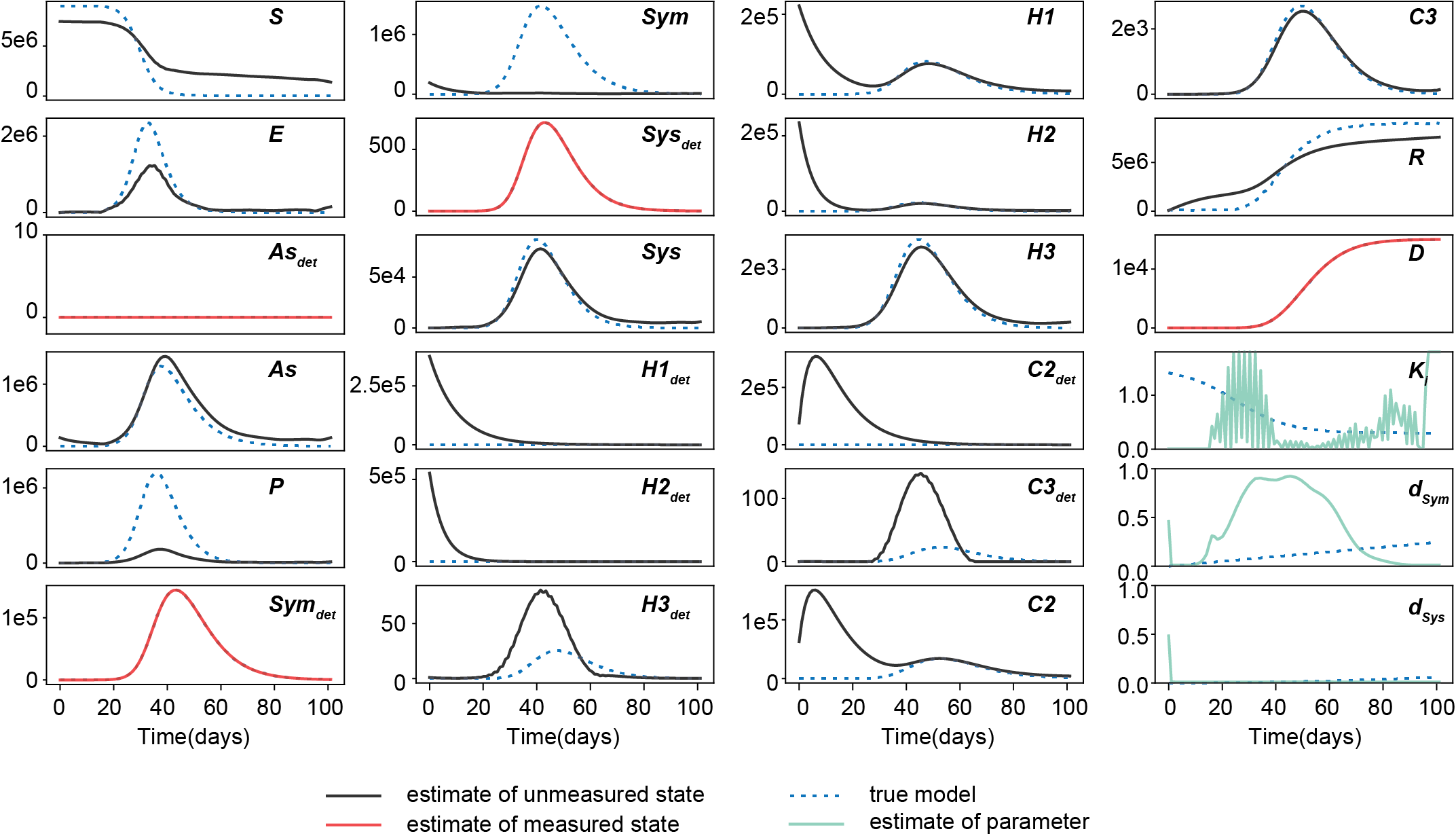}
\caption{\textbf{Estimates for Experiment $\bm{ii}$: without a measurement of Population $\bm{R}$.}  This result is taken at $\beta = 2$, prior to the exponential runaway in the cost.  Estimates of unmeasured states and time-varying parameters are poor. }
 \label{fig:state_nM5_D101_noNoise}
\end{figure}

Indeed, Figure~\ref{fig:state_nM5_D101_noNoise} shows the estimation, taken at $\beta=2$, prior to the runaway behavior.  Note the excellent fit to the measured states and simultaneous poor fit to the unmeasured states.  As no stable solution is found at high $\beta$, we conclude that there exists insufficient information in $As_{det}$, $Sym_{det}$, $Sys_{det}$, and $D$ alone to corral the procedure into a region of state-and-parameter space in which a model solution is possible.  We repeated this experiment with a doubled baseline of 201 days, and noted no improvement (not shown).
\end{multicols}
\begin{multicols}{2}
\end{multicols}
\begin{figure}[H]
  \includegraphics[width=0.9\textwidth]{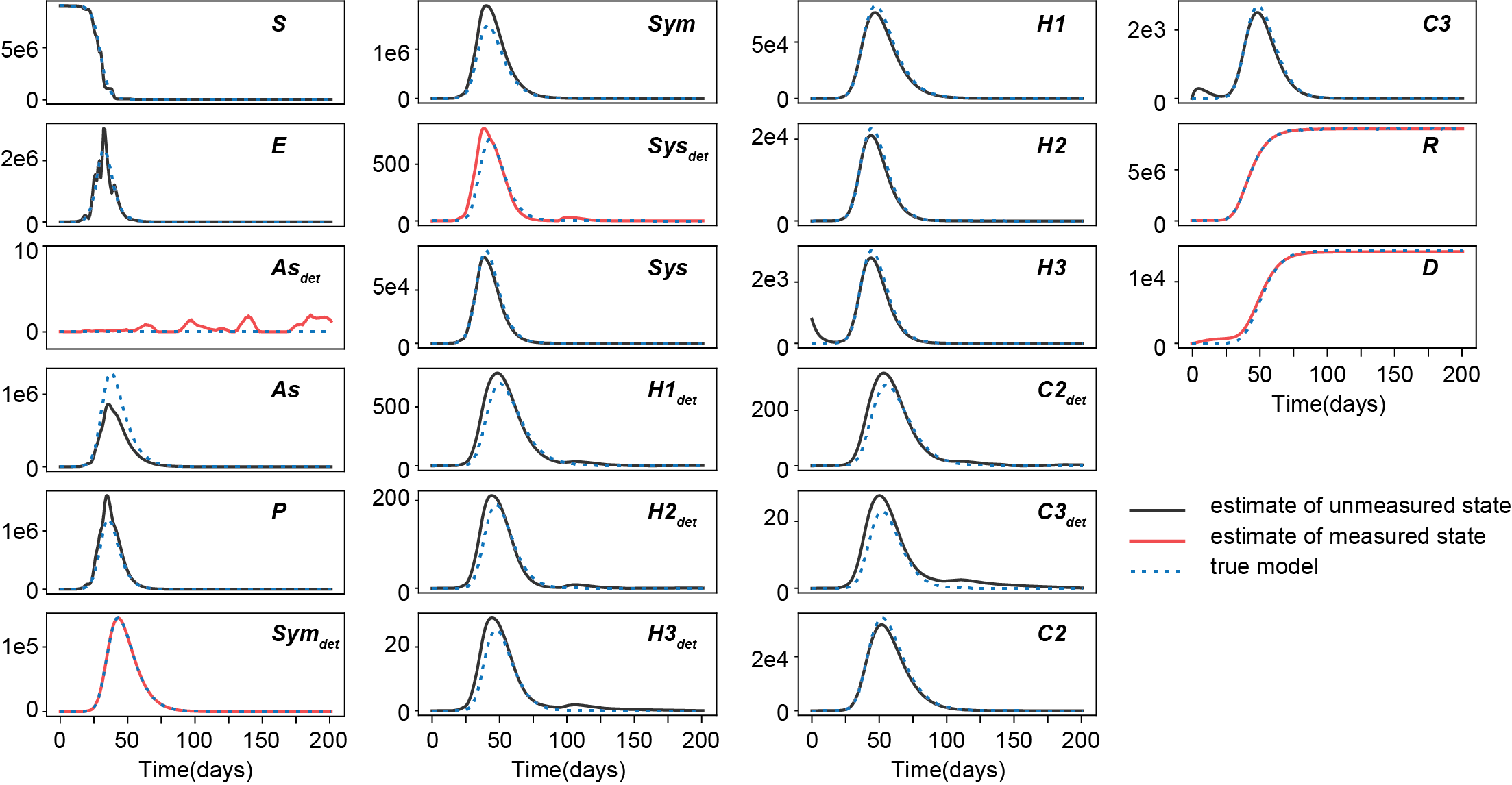}
\caption{\textbf{Estimates for Experiment $\bm{iv}$: low noise added to Population $\bm{R}$ and with a doubled temporal baseline of 201 days.}  The noise added to $R$ propagates to some unmeasured States ($S$, $E$, $As$, and $As_{det}$), but the overall evolution is captured well.  The noise precludes an estimate of the time-varying parameters (not shown).  Results are reported using a value for $\beta$ of 40.}
 \label{fig:state_nM6_D201_Noisy_onlyR}
\end{figure}
\begin{multicols}{2}
\noindent

\subsection{\textbf{Experiments $\mathbf{iii}$ and $\mathbf{iv}$: low noise added}}
In Experiment $iii$, the low noise added to $R$ yielded a poor state and parameter estimate (not shown).  With a doubled temporal baseline of measurements (Experiment $iv$), however, the state estimate became robust to the contamination.  Figure~\ref{fig:state_nM6_D201_Noisy_onlyR} shows this estimate.  While the $\sim$ five percent noise added to Population $R$ propagates to the unmeasured States $S$, $E$, and $P$, the general state evolution is still captured well.  Importantly, the populations entering the hospital are well estimated.  Note that some low state estimates (e.g. $As$) are not perfectly offset by high estimates (e.g. $Sym$).  The addition of noise in these numbers - by definition - breaks the conservation of the population.  Finally, the parameter estimates for Experiment $iv$ do not survive the added contamination (not shown).  

\section{CONCLUSION}
We have endeavoured to illustrate the potential of SDA to systematically identify the specific measurements, temporal baseline of measurements, and degree of measurement accuracy, required to estimate unknown model parameters in a high-dimensional model designed to examine the complex problems that COVID-19 presents to hospitals.  In light of our assumed knowledge of some model parameters, we restrict our conclusions to general comments.  We emphasize that estimation of the full model state requires measurements of the detected cases but not the undetected, provided that the recovered and dead are also measured.  The state evolution is tolerant to low noise in these measurements, while the parameter estimates are not.

The ultimate aim of SDA is to test the validity of model estimation using real data, via \textit{prediction}.  In advance of that step, we are performing a detailed study of the model's sensitivity to contamination in the measurable populations $As_{det}$, $Sym_{det}$, $Sys_{det}$, $R$, and $D$.  Concurrently we are examining means to render the parameter estimation less sensitive to noise, via various additional equality constraints in the cost function, and loosening the assumption of Gaussian-distributed noise.  In particular, we shall require that the time-varying parameters be smoothly-varying.  It will be important to examine the stability of the SDA procedure over a range of choices for parameter values and initial numbers for the infected populations. 

This procedure can be expanded in many directions.  Examples include: 1) defining additional model parameters as unknowns to be estimated, including the fraction of patients hospitalized, the fraction who enter critical care, and the various timescales governing the reaction equations; 2) imposing various constraints regarding the unknown time-varying quantities, particularly transmission rate $K_i$(t), and identifying which forms permit a solution consistent with measurements; 3) examining model sensitivity to the initial numbers within each population; 4) examining model sensitivity to the temporal frequency of data sampling. Moreover, it is our hope that the exercises described in this paper can guide the application of SDA to a host of complicated questions surrounding COVID-19.

\section{ACKNOWLEDGEMENTS}
Thank you to Patrick Clay from the University of Michigan for discussions on inferring exposure rates given social distancing protocols.
\end{multicols}
\section*{Appendix A: Details of the model}
\setlength{\tabcolsep}{5pt}
\begin{table}[H]
\small
\centering
\begin{tabular}{l l} \toprule
 \textit{Variable} & \textit{Description}\\\midrule 
 $S$ & Susceptible \\
 $E$ & Exposed \\
 $As_{det}$ & Asymptomatic, detected \\
 $As$ & Asymptomatic, undetected \\
 $Sym_{det}$ & Symptomatic mild, detected \\
 $Sym$ & Symptomatic mild, undetected \\
 $Sys_{det}$ & Symptomatic severe, detected \\
 $Sys$ & Symptomatic severe, undetected \\
 $H_{1,det}$ & Hospitalized and will recover, detected \\
 $H_{2,det}$ & Hospitalized and will go to critical care and recover, detected \\
 $H_{3,det}$ & Hospitalized and will go to critical care and die, detected \\
 $H_{1}$ & Hospitalized and will recover, undetected \\
 $H_{2}$ & Hospitalized and will go to critical care and recover, undetected \\
 $H_{3}$ & Hospitalized and will go to critical care and die, undetected \\
 $C_{2,det}$ & In critical care and will recover, detected \\
 $C_{3,det}$ & In critical care and will die, detected \\
 $C_{2}$ & In critical care and will recover, undetected \\
 $C_{3}$ & In critical care and will die, undetected \\
 $R$ & Recovered\\
 $D$ & Dead\\\bottomrule
\end{tabular}
\caption{\textbf{State variables of the COVID-19 transmission model.}  The \lq\lq detected\rq\rq\ qualifier signifies that the population has been tested and is positive for COVID-19.} 
\label{tableVariables}
\end{table}
\noindent
\textbf{Reaction equations}\\
\noindent
The blue notation specified by overbrackets denotes the correspondence of specific terms to the reactions between the populations depicted in Figure~\ref{fig:schematic}.
\begin{equation*}\label{eom}
\begin{aligned}
\diff{S}{t} &= - \overbrace{\frac{K_i \cdot S \cdot [infectious + (infectious_{det} \times reduced)]}{N}}^{{\color{blue} \rightarrow E}} \\
& \bullet infectious = As + P + Sym + Sys + H_1 + H_2 + H_3 + C_2 + C_3\\\nonumber
& \bullet infectious_{det} = As_{det} + Sym_{det} + Sys_{det}\\\nonumber
\diff{E}{t} &= \overbrace{K_i \cdot S \cdot [infectious + (infectious_{det} \times reduced)]/N}^{{\color{blue} S \rightarrow}}\\
& - \overbrace{\frac{1 - f_{sympt}}{t_{infection}} \cdot E \cdot d_{As}}^{{\color{blue} \rightarrow As_{det}}} - \overbrace{\frac{1 - f_{sympt}}{t_{infection}} \cdot E \cdot (1.0 - d_{As})}^{{\color{blue} \rightarrow As}} - \overbrace{\frac{f_{sympt}}{t_{infection}} \cdot E}^{{\color{blue} \rightarrow P}}\\\nonumber
\diff{As_{det}}{t} &= \overbrace{\frac{1 - f_{sympt}}{t_{infection}} \cdot E \cdot d_{As}}^{{\color{blue} E \rightarrow}} - \overbrace{\frac{1}{t_{R,a}} \cdot As_{det}}^{{\color{blue} \rightarrow R}}\\
\diff{As}{t} &= \overbrace{\frac{1 - f_{sympt}}{t_{infection}} \cdot E \cdot (1.0 - d_{As})}^{{\color{blue} E \rightarrow}} - \overbrace{\frac{1}{t_{R,a}} \cdot As}^{{\color{blue} \rightarrow R}}
\end{aligned}
\end{equation*}
\begin{equation*}
\begin{aligned}
\diff{P}{t} &= \overbrace{\frac{f_{sympt}}{t_{infection}} \cdot E}^{{\color{blue} E \rightarrow}} - \overbrace{\frac{1-f_{severe}}{t_{sympt}} \cdot P \cdot d_{Sym}}^{{\color{blue} \rightarrow Sym_{det}}} - \overbrace{\frac{1-f_{severe}}{t_{sympt}} \cdot P \cdot (1.0 - d_{Sym})}^{{\color{blue} \rightarrow Sym}} - \overbrace{\frac{f_{severe}}{t_{sympt}} \cdot P \cdot d_{Sys}}^{{\color{blue} \rightarrow Sys_{det}}} - \overbrace{\frac{f_{severe}}{t_{sympt}} \cdot P \cdot (1.0 - d_{Sys})}^{{\color{blue} \rightarrow Sys}}\\
\diff{Sym_{det}}{t} &= \overbrace{\frac{1-f_{severe}}{t_{sympt}} \cdot P \cdot d_{Sym}}^{{\color{blue} P \rightarrow}} - \overbrace{\frac{1}{t_{R,m}} \cdot Sym_{det}}^{{\color{blue} \rightarrow R}}\\
\diff{Sym}{t} &= \overbrace{\frac{1-f_{severe}}{t_{sympt}} \cdot P \cdot (1.0 - d_{Sym})}^{{\color{blue} P \rightarrow}} - \overbrace{\frac{1}{t_{R,m}} \cdot Sym}^{{\color{blue} \rightarrow R}}\\
\diff{Sys_{det}}{t} &= \overbrace{\frac{f_{severe}}{t_{sympt}} \cdot P \cdot d_{Sys}}^{{\color{blue} P \rightarrow}} - \overbrace{\frac{f_H}{t_H} \cdot Sys_{det}}^{{\color{blue} \rightarrow H_{{1,det}}}} - \overbrace{\frac{f_C}{t_H} \cdot Sys_{det}}^{{\color{blue} \rightarrow H_{{2,det}}}} - \overbrace{\frac{f_D}{t_H} \cdot Sys_{det}}^{{\color{blue} \rightarrow  H_{{3,det}}}}\\
\diff{Sys}{t} &= \overbrace{\frac{f_{severe}}{t_{sympt}} \cdot P \cdot (1.0 - d_{Sys})}^{{\color{blue} P \rightarrow}} - \overbrace{\frac{f_H}{t_H} \cdot Sys}^{{\color{blue} \rightarrow H_1}} - \overbrace{\frac{f_C}{t_H} \cdot Sys}^{{\color{blue} \rightarrow H_2}} - \overbrace{\frac{f_D}{t_H} \cdot Sys}^{{\color{blue} \rightarrow H_3}}\\
\diff{H_{1,det}}{t} &= \overbrace{\frac{f_H}{t_H} \cdot Sys_{det}}^{{\color{blue} Sys_{det} \rightarrow}} - \overbrace{\frac{1}{t_{R,h}} \cdot H_{1,det}}^{{\color{blue} \rightarrow R}}\\
\diff{H_{2,det}}{t} &= \overbrace{\frac{f_C}{t_H} \cdot Sys_{det}}^{{\color{blue} Sys_{det} \rightarrow}} - \overbrace{\frac{1}{t_C} \cdot H_{2,det}}^{{\color{blue} \rightarrow C_{2,det}}}\\
\diff{H_{3,det}}{t} &= \overbrace{\frac{f_D}{t_H} \cdot Sys_{det}}^{{\color{blue} Sys_{det} \rightarrow}} - \overbrace{\frac{1}{t_C} \cdot H_{3,det}}^{{\color{blue} \rightarrow C_{3,det}}}\\
\diff{H_1}{t} &= \overbrace{\frac{f_H}{t_H} \cdot Sys}^{{\color{blue} Sys \rightarrow}} - \overbrace{\frac{1}{t_{R,h}} \cdot H_1}^{{\color{blue} \rightarrow R}}\\
\diff{H_2}{t} &= \overbrace{\frac{f_C}{t_H} \cdot Sys}^{{\color{blue} Sys \rightarrow}} - \overbrace{\frac{1}{t_C} \cdot H_2}^{{\color{blue} \rightarrow C_2}}\\
\diff{H_3}{t} &= \overbrace{\frac{f_D}{t_H} \cdot Sys}^{{\color{blue} Sys \rightarrow}} - \overbrace{\frac{1}{t_C} \cdot H_3}^{{\color{blue} \rightarrow C_3}}\\
\diff{C_{2,det}}{t} &= \overbrace{\frac{1}{t_C} \cdot H_{2,det}}^{{\color{blue} H_{2,det} \rightarrow}} - \overbrace{\frac{1}{t_{R,c}} \cdot C_{2det}}^{{\color{blue} \rightarrow R}}\\
\diff{C_{3,det}}{t} &= \overbrace{\frac{1}{t_C} \cdot H_{3,det}}^{{\color{blue} H_{3,det} \rightarrow}} - \overbrace{\frac{1}{t_D} \cdot C_{3,det}}^{{\color{blue} \rightarrow D}}\\
\diff{C_2}{t} &= \overbrace{\frac{1}{t_C} \cdot H_2}^{{\color{blue} H_2 \rightarrow}} - \overbrace{\frac{1}{t_{R,c}} \cdot C_2}^{{\color{blue} \rightarrow R}}\\
\diff{C_3}{t} &= \overbrace{\frac{1}{t_C} \cdot H_3}^{{\color{blue} H_3 \rightarrow}} - \overbrace{\frac{1}{t_D} \cdot C_3}^{{\color{blue} \rightarrow D}}\\
\end{aligned}
\end{equation*}
\begin{equation*}
\begin{aligned}
\diff{R}{t} &= \overbrace{\frac{1}{t_{R,a}} \cdot As_{det}}^{{\color{blue} As_{det} \rightarrow}} + \overbrace{\frac{1}{t_{R,a}} \cdot As}^{{\color{blue} As \rightarrow}} + \overbrace{\frac{1}{t_{R,m}} \cdot Sym_{det}}^{{\color{blue} Sym_{det} \rightarrow}} + \overbrace{\frac{1}{t_{R,m}} \cdot Sym}^{{\color{blue} Sym \rightarrow}} + \overbrace{\frac{1}{t_{R,h}} \cdot H_{1,det}}^{{\color{blue} H_{1,det} \rightarrow}} + \overbrace{\frac{1}{t_{R,h}} \cdot H_1}^{{\color{blue} H_1 \rightarrow}} + \overbrace{\frac{1}{t_{R,c}} \cdot C_{2,det}}^{{\color{blue} C_{2,det} \rightarrow}} + \overbrace{\frac{1}{t_{R,c}} \cdot C_2}^{{\color{blue} C_2 \rightarrow}}\\
\diff{D}{t} &= \overbrace{\frac{1}{t_D} \cdot C_{3,det}}^{{\color{blue} C_{3,det} \rightarrow}} + \overbrace{\frac{1}{t_D} \cdot C_3}^{{\color{blue} C_3 \rightarrow}} 
\end{aligned}
\end{equation*}
\begin{multicols}{2}

\end{multicols}

\setlength{\tabcolsep}{5pt}
\begin{table}[H]
\small
\centering
\begin{tabular}{p{2cm}|p{11cm}|p{3cm}}\toprule
 \textit{Parameter} & \textit{Description} & \textit{Value} \\\midrule
  $N$ & Total population & 9,000,000\\
  $reduced$ & The property that a detected case is likely to transmit less, via successful quarantine) &  0.2\\
  $\bm{K_i(t)}$ & \textbf{Transmission rate} & See \textit{Appendix B}\\
 $d_{As}(t)$ & Detection probability of asymptomatic cases & 0.0\\
 $\bm{f_{sympt}}$ & \textbf{Fraction of positive cases that produce symptoms} & \textbf{0.6}~\cite{fSympt}\\
 $t_{infection}$ & Time from exposure to infection & 4.0~\cite{li_substantial_2020} \\
 $t_{R,a}$ & Time to recovery for asymptomatics & 8.0 \textit{Assumed to be same as } $t_{R,m}$ \\
 $\bm{d_{Sym}(t)}$ & \textbf{Detection probability of mild symptomatics} & See \textit{Appendix B}\\
 $\bm{d_{Sys}(t)}$ & \textbf{Detection probability of severe symptomatics} & See \textit{Appendix B}\\
 $\bm{f_{severe}}$ & \textbf{Fraction of symptomatics that are severe} & \textbf{0.07}~\cite{salje2020estimating}\\
 $t_{sympt}$ & Time to symptoms, for symptomatics & 4.0~\cite{jing_estimation_2020, li_substantial_2020} \\
 $t_{R,m}$ & Time from symptoms to recovery, for mild symptomatics & 8.0~\cite{wolfel_virological_2020}\footnote{As described in ~\cite{wolfel_virological_2020}, viral load can be high and detectable for up to 20 days. We choose a shorter duration of infectiousness to capture the time during which transmissibility is highest.} \\
 $f_{H}$ & Fraction of severe cases that are hospitalized and then recover: $f_{H} = 1.0 - f_C - f_D$ & 0.66\\
 $f_{C}$ & Fraction of severe cases that require critical care and then recover & 0.3~\cite{lewnard2020incidence}\\
 $f_{D}$ & Fraction of severe cases that die & 0.04~\cite{wang_clinical_2020}\\
 $t_H$ & Time from symptoms to hospital, for severe symptomatics & 5.0~\cite{huang2020clinical} \\
 $t_{R,h}$ & Time from entering hospital to recovery, for severe symptomatics that do not require critical care & 10.0~\cite{wang_clinical_2020, lewnard2020incidence}\\
 $t_C$ & Time from entering hospital to critical care, for severe symptomatics & 5.0~\cite{huang2020clinical} \\
 $t_{R,c}$ & Time from entering critical care to recovery for severe symptomatics & 10.0~\cite{bi2020epidemiology}\\
 $t_D$ & Time from entering critical care to death, for severe symptomatics & 5.0~\cite{yang2020clinical} \\\bottomrule
 \end{tabular}
\caption{\textbf{The model parameters, with the unknown parameters to be estimated denoted in boldface.}  The unknown parameters $K_i$, $_{Sym}$, and $d_{Sys}$ are taken to be time-varying.  The unknown parameters $f_{sympt}$ and $f_{severe}$ are taken to be intrinsic properties of the disease and therefore constant numbers.  The detection probability of asymptomatic cases is taken to be known and zero.  Units of time are days.} 
\label{tableKnown}
\end{table}
\section*{Appendix B: Unknown time-varying parameters to be estimated}
The unknown parameters assumed to be time-varying are the transmission rate $K_i$, and the detection probabilities $d_{Sym}$ and $d_{Sys}$ for mild and severe symptomatic cases, respectively.

The transmission rate in a given population for a given infectious disease is measured in effective contacts per unit time.  This may be expressed as the total contact rate (the total number of contacts, effective or not, per unit time), multiplied by the risk of infection, given contact between an infectious and a susceptible individual.  The total contact rate can be impacted by social behavior.

In this first employment of SDA upon a pandemic model of such high dimensionality, we chose to represent $K_i$ as a relatively constant value that undergoes one rapid transition corresponding to  a single social distancing mandate.  As noted in \textit{Experiments}, social distancing rules were imposed in New York City roughly 25 days following the first reported case.  We thus chose $K_i$ to transition between two relatively constant levels, roughly 25 days following time $t_0$.  Specifically, we wrote $K_i$(t) as:
\begin{align*}\label{eq:timeVaryingPs}
  Ki(t) &= -f \cdot \frac{1}{e^{(T - t)/s} + 1} + \xi.
\end{align*}
\noindent
The parameter T was set to 25, beginning four days prior to the first report of a detection in NYC~\cite{firstInNYC} to the imposition of a stay-home order in NYC on March 22~\cite{PAUSEorder}.  The parameter s governs the steepness of the transformation, and was set to 10.  Parameters $f$ and $\xi$ were then adjusted to 1.2 and 1.5, to achieve a transition from about 1.4 to 0.3.

For detection probabilities $d_{Sym}$ and $d_{Sys}$, a linear and quadratic form, respectively, were chosen to preclude symmetries, and both were optimistically taken to increase with time:
\begin{align*}
  dSym(t) &= 0.2 \cdot t\nonumber\\
  dSys(t) &= 0.1 \cdot t^2\nonumber
\end{align*}
\noindent
Finally, each time series was normalized to the range: [0:1], via division by their respective maximum values.

\section*{Appendix C: Technical details of the inference experiments}
The simulated data were generated by integrating the reaction equations (\textit{Appendix A}) via a fourth-order adaptive Runge-Kutta method encoded in the Python package odeINT.  A step size of one (day) was used to record the output.  Except for the one instance noted in \textit{Results} regarding Experiment $i$, we did not examine the sensitivity of estimations to the temporal sparsity of measurements.  The initial conditions on the populations were: $S_0 = N - 1$ (where $N$ is the total population), $As_0 = 1$, and zero for all others. 

For the noise experiments, the noise added to the simulated $Sym_{det}$, $Sys_{det}$, and $R$ data were generated by Python's \textit{numpy.random.normal} package, which defines a normal distribution of noise.  For the \lq\lq low-noise\rq\rq\ experiments, we set the standard deviation to be the respective mean of each distribution, divided by 100.  For the experiments using higher noise, we multiplied that original level by a factor of ten.  For each noisy data set, the absolute value of the minimum was then added to each data point, so that the population did not drop below zero.

The optimization was performed via the open-source Interior-point Optimizer (Ipopt)~\cite{wachter2009short}.  Ipopt uses a Simpson’s rule method of finite differences to discretize the state space, a Newton’s method to search, and a barrier method to impose user-defined bounds that are placed upon the searches.  We note that Ipopt's search algorithm treats state variables as independent quantities, which is not the case for a model involving a closed population.  This feature did not affect the results of this paper.  Those interested in expanding the use of this tool, however, might keep in mind this feature.  One might negate undesired effects by, for example, imposing equality constraints into the cost function that enforce the conservation of $N$.

Within the annealing procedure described in \textit{Methods}, the parameter $\alpha$ was set to 2.0, and $\beta$ ran from 0 to 38 in increments of 1.  The inverse covariance matrix for measurement error ($R_m$) was set to 1.0, and the initial value of the inverse covariance matrix for model error ($R_{f,0}$) was set to $10^{-7}$. 

For each of the four simulated experiments, twenty paths were searched, beginning at randomly-generated initial conditions for parameters and state variables.  All simulations were run on a 720-core, 1440-GB, 64-bit CPU cluster.
\bibliographystyle{unsrt}
\bibliography{bib_edited}
\end{document}